\documentstyle[epsfig,12pt]{article} \textwidth=16cm
\begin{document}
\title {Temperature Equilibration Rate of Quasi-Monoenergetic Deuteron Beam in a Fusion Plasmas}
\author{M. Mahdavi$^{1}$\thanks{email:
        {\tt m.mahdavi@umz.ac.ir }}$~ $ R. Azadifar $^{1}$\thanks{email:
        {\tt r.azadifarr@gmail.com}},$~ $ and T. Khoorokhi$^{2}$\thanks{email:
        {\tt t.koohrokhi@gu.ac.ir }} }

\date{{\small $^{1}$Physics Department, University of Mazandaran, P. O. Box
47415-416, Babolsar, Iran\\}{\small $^{2}$Physics Department,
Faculty of Sciences, Golestan University, Shahid Beheshti Street,
P.O. Box 155, Gorgan, Iran\\}}

\maketitle
\newpage


\begin{abstract}
Thermal equilibrium rate can play an important role in the energy
deposition of beam to the fuel in fast ignition due to high
temperature difference between projectile ions and background plasma
ions. In this study the temperature equilibration rate of a
quasi-monoenergetic deuteron beam with an equimolar
Deuterium-Tritium fusion plasma with a Maxwellian energy
distribution is calculated by kinetic theory equations. In this
theory, binary collisions is described by the Boltzman equation and
collective effects is described by the Lenard-Balescu equation. The
obtained results show that at higher background temperatures,
$T_{b}=100 keV$, the ions interactions effect in the temperature
equilibration rate increases because the deuteron beam exchanges most of
its energy with ions plasma.\\

 \end{abstract}
\newpage


\section{Introduction}
A typical plasma which is formed at ignition and burn stages of an
inertial confinement fusion fuel is known as a hot dense plasma [1].
The plasma temperature (T$\simeq$ 10-100 keV) exceeds the sun's
temperature during burning and the capsule is compressed to high
densities ($\rho \simeq 300-500~\textrm{g}\textrm{cm}^{-3})$ at the
ignition instance [2]. The determinative conditions of a plasma are
determined by it's temperature and density, spontaneously [3].
Despite of high temperature and density, the fusion plasma in the
inertial confinement fuel capsule is a weakly coupled plasma [4]. A
non-interacting weakly coupled plasma is described by
Maxwell-Boltzmann distribution function at the thermal equilibrium
[5]. Nevertheless, a fusion plasma is an interacting plasma whose
temperature of plasma species (different ions and electrons) is
changed by the energy gain and loss mechanisms. For different
characteristic properties of particles (mass, charge etc), the
plasma species have different temperatures during burning of fuel.
The particles with different temperatures exchange their energy
together via collisions and collective interactions [6]. The
dynamical analyses of the igniting and burning of fusion plasma
require exact calculations
for temperature equilibration rate of plasma species.\\
The plasma kinetic theory is derived from statistical mechanics and
describes the evolution of particles distribution functions [7]. In
the plasma kinetic theory, the Boltzmann equation for Coulomb and
nuclear elastic scattering describes the short-distance, hard
collisions of the plasma particles whereas the long-distance,
collective excitations of the plasma, is described by the
Lenard-Balescu equation [8,9]. This theory has been used for
calculation of temperature equilibration rate of two particles with
Maxwell-Boltzmann distribution functions at two different
temperatures. On the other hand, ion fast ignition using
quasi-monoenergetic ion beams as ignitor have been proposed and
studied after production of laser accelerated quasi-monoenergetic
ion beams, experimentally [10,11]. The narrow energy spread and high
conversion efficiency of quasi-monoenergetic laser-driven
high-current ions, make them very suitable for local energy
deposition. Among the ion beams which were studied as ignitor, the
deuteron beam has special privilege, because the deuteron beam can
experience fusion reactions during deposition of its energy via
stopping into the fuel [12,13]. These reasons promoted us to study a
quasi-monoenergetic deuteron beam as the projectile that deposits
its energy into an equimolar deuterium-tritium plasma.
The obtained results are remarkable and applicable for accurate simulations and advanced codes. \\


 \section{Projectile and Background Particles Distribution Functions}

The distribution function of quasi-monoenergetic projectile beam has
been considered as the Gaussian function as [14],
\begin{equation}
f_{p}\left(E_{p}\right)=\frac{n_{p}\sqrt{\alpha}}{\Delta\sqrt{\pi}}
\textrm{exp}\left[-\alpha\left(\frac{E_{p}-E_{0}}{\Delta}\right)^{2}\right].
\end{equation}
where $n_{p}$ is number of densities of projectile ions and
$\alpha=4\textrm{ln}(2)$. The $E_{0}=<E_{p}>$ is the average kinetic
energy of ions in which for quasi-monoenergetic distribution
function is equal to projectile temperature $T_{p}$. The quantity
$\Delta = \Delta E/E$ is the energy spread so that high quality
projectile beam has $10 \%$ [15]. If the energy spread $\Delta$
(full width at half maximum, FWHM)is $10 \%$, this distribution
refer to as "quasi-monoenergetic". Corresponding the relation $\int
f_{p}\left(\textbf{v}_{p}\right)d^{3}v_{p}=\int
f_{p}\left(E_{p}\right)dE_{p}$ and projectile kinetic energy
$E_{p}=\frac{1}{2}m_{p}v^{2}_{p}$, the quasi-monoenergetic velocity
distribution in the phase space is,
\begin{equation}
f_{p}\left(\textbf{v}_{p}\right)=\frac{2n_{p}(\pi
\hbar^{2})^{3/2}\sqrt{\alpha}}{\Delta m^{2}_{p}v_{p}}
\textrm{exp}\left[-\alpha\left(\frac{m_{p}v^{2}_{p}}{2\Delta}-\frac{E_{0}}{\Delta}\right)^{2}\right].
\end{equation}
where $m_{p}$ and $v_{p}$ are the mass and velocity of projectile,
respectively. In a fusion plasma, the velocity distribution function
of background particles $b$ at thermal equilibrium is considered as
Maxwellian form,
\begin{equation}
f_{b}\left(\textbf{v}_{b}\right)=n_{b}\left(\frac{2\pi\hbar^{2}\beta_{b}}{m_{b}}\right)^{3/2}
\textrm{exp}\left(-\frac{1}{2}\beta_{b}m_{b}v^{2}_{b}\right).
\end{equation}
where $\beta_{b}=1/T_{b}$, $m_{b}$, $v_{b}$ and $n_{b}$ are the
inverse temperature, mass, velocity and number density of 'b'
particle, respectively. For next purposes, we need multiply the two
projectile and background distribution functions,
\begin{equation}
 f_{p}\left(\textbf{v}_{p}\right)f_{b}\left(\textbf{v}_{b}\right)=
 \frac{n_{p}n_{b}2^{5/2}(\pi \hbar^{2})^{3}\sqrt{\alpha}}{\Delta m^{2}_{p}m^{3/2}_{b}v_{p}T^{3/2}_{b}}
\textrm{exp}\left[-\alpha\left(\frac{m_{p}v^{2}_{p}}{2\Delta}\frac{T_{p}}{\Delta}\right)^{2}-\frac{1}{2}\beta_{b}m_{b}v^{2}_{b}\right].
\end{equation}
Furthermore, as we will see soon, it is useful to change laboratory
velocities to the
 center of mass $\textbf{V}_{c}$ and relative
 $\textbf{v}_{pb}=\textbf{v}_{p}-\textbf{v}_{b}$ velocities as,
\begin{eqnarray}
\left\{ \begin{array}{ll}
                    \textbf{v}_{p}=\textbf{V}_{c}+\frac{m_{b}}{M_{pb}}\textbf{v}_{pb}\Rightarrow
                    v^{2}_{p}=V^{2}_{c}+\frac{m^{2}_{b}}{M^{2}_{pb}}v^{2}_{pb}+2\textbf{V}_{C}~\textbf{.}~\textbf{v}_{pb}\frac{m_{b}}{M_{pb}} \\
                    \textbf{v}_{b}=\textbf{V}_{b}-\frac{m_{p}}{M_{pb}}\textbf{v}_{pb}\Rightarrow
                    v^{2}_{b}=V^{2}_{c}+\frac{m^{2}_{p}}{M^{2}_{pb}}v^{2}_{pb}-2\textbf{V}_{C}~\textbf{.}~\textbf{v}_{pb}\frac{m_{p}}{M_{pb}},
                    \end{array}
                    \right.
\end{eqnarray}
Replacing these translations into Eq. (4), leads,
\begin{eqnarray}
&&f_{p}\left(\textbf{v}_{p}\right)f_{b}\left(\textbf{v}_{b}\right)=
 \frac{n_{p}n_{b}2^{5/2}(\pi \hbar^{2})^{3}\sqrt{\alpha}}{\Delta m^{2}_{p}m^{3/2}_{b}v_{p}T^{3/2}_{b}}
 \left(V^{2}_{c}+\frac{m^{2}_{b}}{M^{2}_{pb}}v^{2}_{pb}+2\textbf{v}_{pb}~\textbf{.}~\textbf{V}_{c}\frac{m_{b}}{M_{pb}}\right)^{-1/2}\times \nonumber \\
&&
\textrm{exp}\left\{-\frac{\alpha}{\Delta^{2}}\left[\frac{m^{2}_{p}V^{4}_{c}}{4}
+\frac{m^{4}_{pb}v^{4}_{pb}}{4m^{2}_{p}}+m^{2}_{pb}(\textbf{v}_{pb}.\textbf{V}_{c})^{2}-
\left(m_{p}T_{p}-\frac{\Delta^{2}m_{b}}{2\alpha T_{b}}-m_{pb}m_{p}\textbf{v}_{pb}.\textbf{V}_{c}\right)V^{2}_{c} \right. \right. \nonumber\\
&& \left. \left.
-\left(\frac{T_{p}}{m_{p}}-\frac{\Delta^{2}}{2m_{b}\alpha
T_{b}}-\frac{m_{pb}}{m_{p}}\textbf{v}_{pb}.\textbf{V}_{c}
-\frac{V^{2}_{c}}{2}\right)m^{2}_{pb}v^{2}_{pb}-\left(2T_{p}+\frac{\Delta^{2}}{\alpha
T_{b}}\right)m_{pb}\textbf{v}_{pb}.\textbf{V}_{c}+T^{2}_{p}\right]\right\}
\end{eqnarray}
This relation will be used in the next sections.

 \section{The Energy Exchange Rate Due to Binary Collisions}
The energy exchange rate between projectile ions and background
particles of plasma due to binary collisions is evaluated by
Boltzmann equation [16]:
\begin{equation}
\frac{d\varepsilon_{pb}}{dt}=\int\frac{d^{3}\textbf{p}_{b}}{(2\pi\hbar)^{3}}\frac{d^{3}\textbf{p}_{p}}{(2\pi\hbar)^{3}}
f_{b}\left(\textbf{p}_{b}\right)f_{p}\left(\textbf{p}_{p}\right)v_{pb}\int
d\Omega\left(\frac{d\sigma_{pb}}{d\Omega}\right)\left[E'_{p}-E_{p}\right],
\end{equation}
where $E_{p}$ and $E'_{p}$ are the kinetic energies of projectile
before and after one collision, respectively. Since in an elastic
collision, Coulomb as well as nuclear potentials are contributed to
the scattering, the total elastic scattering cross section
$(d\sigma_{pb}/d\Omega)$ is the sum of Coulomb
$(d\sigma^{\textrm{Coul}}_{pb}/d\Omega)$ and nuclear
$(d\sigma^{\textrm{NI}}_{pb}/d\Omega)$ cross sections, [17]. The
experimental data for these quantities are available in Ref. [18].
In the quantum mechanics, there are differences between the
scattering of identical particles, such as deuterium-deuterium
(D+D), and distinguishable particles such as deuterium-tritium
(D+T). The Coulomb scattering is represented by the Rutherford
formula for distinguishable particles [19],
\begin{equation}
\left(\frac{d\sigma}{d\Omega}\right)_{cd}^{\textrm{Coul}}=\frac{\eta^{2}}{k^{2}(1-\mu)^{2}},
\end{equation}
and for identical particles is presented as,
\begin{equation}
\left(\frac{d\sigma}{d\Omega}\right)_{ci}^{\textrm{Coul}}=\frac{2\eta^{2}}{k^{2}(1-\mu^{2})}
\left[\frac{1+\mu^{2}}{1-\mu^{2}}+\frac{(-1)^{2s}}{2s+1}
\cos\left(\eta \ln\frac{1+\mu}{1-\mu}\right)\right],
\end{equation}
where $s$ is the spin of identical particles, $\mu=\cos\Theta$ is
cosine of the scattering angle in the center-of-mass system, $k$ is
particle wave number and $\eta$ is the dimensionless Coulomb
parameter,
\begin{equation}
\eta_{pb}=\frac{Z_{p}Z_{b}e^{2}}{\hbar v_{pb}},
\end{equation}
The nuclear elastic scattering cross section for identical particles
may be written as,
\begin{eqnarray}
\left(\frac{d\sigma}{d\Omega}\right)_{ci}^{\textrm{NI}}&=&
-\frac{2\eta}{1-\mu^{2}}\textrm{Re}\left\{\sum_{\ell=0}^{NL}\left[\left(1+\mu\right)
\exp\left(i\eta\ln\frac{1-\mu}{2}\right)+(-1)^{\ell}\left(1-\mu\right) \right. \right. \nonumber\\
&& \left. \left. \times
\exp\left(i\eta\ln\frac{1+\mu}{2}\right)\right]\frac{2\ell+1}{2}a_{\ell}(E)P_{\ell}(\mu)
\right\}+\sum_{\ell=0}^{NL}\frac{4\ell+1}{2}b_{\ell}(E)P_{2\ell}(\mu),
\end{eqnarray}
and the nuclear cross section for distinguishable particles is,
\begin{eqnarray}
\left(\frac{d\sigma}{d\Omega}\right)_{cd}^{\textrm{NI}}&=&
-\frac{2\eta}{1-\mu}\textrm{Re}\left\{\exp\left(i\eta\ln\frac{1-\mu}{2}\right)
\sum_{\ell=0}^{NL}\frac{2\ell+1}{2}a_{\ell}(E)P_{\ell}(\mu)
\right\}\nonumber\\
&&+\sum_{\ell=0}^{2NL}\frac{2\ell+1}{2}b_{\ell}(E)P_{\ell}(\mu),
\end{eqnarray}
The value of $\textrm{NL}$ represents the highest partial wave
contributing to nuclear scattering. $a_{\ell}$ are complex
coefficients for expanding the trace of the nuclear scattering
amplitude matrix and $b_{\ell}$ coefficients are real coefficients
for expanding the nuclear scattering cross section which are derived
from experimental data. $P_{\ell}(\mu)$ and $P_{2\ell}(\mu)$ are
also Legendre functions.

The change in projectile kinetic energy as a result of the collision
is equal to [20],
\begin{equation}
  E'_{p}-E_{p}= m_{pb}V_{c}v_{pb} \left(x(\mu-1)+\sqrt{(1-x^{2})(1-\mu^{2})}\right),
\end{equation}
where $\cos\varphi=x$ is the angle cosine between relative
$\textbf{v}_{pb}$ and center of mass $\textbf{V}_{c}$ velocities and
$m_{pb}=m_{p}m_{b}/(m_{p}+m_{b})$ is the reduced mass of projectile
and background particles. Since the scattering in the C.M. frame is
axially symmetric about the relative speed $\textbf{v}_{pb}$,
  transverse components average is zero in the scattering process,
 so the second term on the right of this equation will be
 removed.

Since there is a unit Jacobian
($j(\textbf{v}_{p},\textbf{v}_{b};\textbf{V}_{c},\textbf{v}_{pb})=1$)
in passing to center of mass coordinate
 $d\textbf{v}_{p}d\textbf{v}_{b}=j(\textbf{v}_{p},\textbf{v}_{b};\textbf{V}_{c},\textbf{v}_{pb})d\textbf{V}_{c}d\textbf{v}_{pb}$, by replacing the Eqs. (6) and (13) into Eq. (7) and changing variables, we have,
\begin{eqnarray}
&&\frac{d\varepsilon^{B}_{pb}}{dt}=
 \frac{n_{p}n_{b}m_{p}m^{3/2}_{b}\sqrt{2\alpha}}{\Delta T^{3/2}_{b}}
\int^{\infty}_{0}dV_{c}\int^{\infty}_{0}dv_{pb} \int^{1}_{-1}xdx \left(V^{2}_{c}+\frac{m^{2}_{b}v^{2}_{pb}}{M^{2}_{pb}}+\frac{2m_{b}v_{pb}V_{c}x}{M_{pb}}\right)^{-1/2}\times \nonumber \\
&&
\textrm{exp}\left\{-\frac{\alpha}{\Delta^{2}}\left[\frac{m^{2}_{p}V^{4}_{c}}{4}
+\frac{m^{4}_{pb}v^{4}_{pb}}{4m^{2}_{p}}+m^{2}_{pb}(v_{pb}V_{c}x)^{2}-
\left(m_{p}T_{p}-\frac{\Delta^{2}m_{b}}{2\alpha T_{b}}-m_{pb}m_{p}v_{pb}V_{c}x\right)V^{2}_{c} \right. \right. \nonumber\\
&& \left. \left.
-\left(\frac{T_{p}}{m_{p}}-\frac{\Delta^{2}}{2m_{b}\alpha
T_{b}}-\frac{m_{pb}}{m_{p}}v_{pb}V_{c}x
-\frac{V^{2}_{c}}{2}\right)m^{2}_{pb}v^{2}_{pb}-\left(2T_{p}+\frac{\Delta^{2}}{\alpha T_{b}}\right)m_{pb}v_{pb}V_{c}x+T^{2}_{p}\right]\right\} \nonumber\\
&&I_{1}(E_{c}),
\end{eqnarray}
where $M_{pb}=m_{p}+m_{b}$ is the total mass and the integral
$I_{1}(E_{c})$ is defined as,
\begin{equation}
I_{1}(E_{c})=\int_{\mu_{min}}^{1}\left(\frac{d\sigma_{pb}}{d\Omega}\right)\left(\mu-1\right)
d\mu,
\end{equation}
In quantum mechanics between the particle that scatters at an angle
$\Theta$ from the one that scatters at $(\pi-\Theta)$ for identical
particles are not distinguishable. As a result, $\mu_{min}$ is zero
for identical particles and it is one for distinguishable particles.
The integral $I_{1}(E_{c})$ is calculated by extracting experimental
data for elastic scattering cross sections. The results are shown in
Fig. (1) for D+D, and in Fig. (2) for D+T scattering, respectively.

Changing variable $v_{pb}$ to the total kinetic energy of particles
at the center of mass system $E_{c}=\frac{1}{2}m_{pb}v^{2}_{pb}$,
the Eq. (14) may be written as the simple form,
\begin{equation}
\frac{d\varepsilon^{B}_{pb}}{dt}=
 \frac{4n_{p}n_{b}m_{p}m^{3/2}_{b}\sqrt{\alpha}}{\Delta m^{3/2}_{pb} T^{3/2}_{b}}
\int^{\infty}_{0}dE_{c}E^{3/2}_{c}I_{1}(E_{c})\int^{\infty}_{0}dV_{c}V^{3}_{c}I_{2}(E_{c},V_{c}),
\end{equation}
where the integral $I_{2}(E_{c})$ is defined as,
\begin{equation}
I_{2}(E_{c},V_{c})=\int_{-1}^{1}dx
x\left(V^{2}_{c}+\frac{2E_{c}}{m^{2}_{p}}+\frac{V_{c}x(8E_{c}m_{pb})^{1/2}}{m_{p}}\right)^{-1/2}
\textrm{exp}\left(-ax^2+bx+c\right)
\end{equation}
where the coefficients are,
\begin{eqnarray}
\left\{ \begin{array}{ll}
a=\frac{2\alpha}{\Delta^{2}}m_{pb}V^{2}_{c}E_{c}\\
b=\frac{\alpha}{\Delta^{2}}\left(2T_{p}+\frac{\Delta^{2}}{\alpha T_{b}}-m_{p}V^{2}_{c}-\frac{2E_{c}m_{pb}}{m_{p}}\right)V_{c}(2E_{c}m_{pb})^{1/2}\\
c=\frac{\alpha}{\Delta^{2}}\left(m_{p}T_{p}V^{2}_{c}+\frac{2E_{c}m_{pb}T_{p}}{m_{p}}-T^{2}_{p}
-\frac{m^{2}_{pb}E^{2}_{c}}{m^{2}_{p}}-m_{pb}V^{2}_{c}E_{c}\right)-\frac{m_{b}V^{2}_{c}}{2T_{b}}-\frac{m_{pb}E_{c}}{T_{b}m_{b}},
                    \end{array}
                    \right.
\end{eqnarray}
The temperature equilibration rate due to binary collisions
(Eq.(16)) versus temperature difference $\Delta T=T_{p}-T_{b}$
between projectile and background particles for background
temperatures $T_{b}$=1 keV, 10 keV and 100 keV are depicted in Figs.
(3), (4) and (5), respectively.
 The highest projectile temperature is chosen $T_{D}$=10 MeV and the background density is taken $\rho_{b}=300
~\textrm{gcm}^{-3}$. In these figures, the ions and electron
contributions denoted by the dashed and dotted lines, respectively,
and the solid line represents the sum of ions and electron
contributions. The results predicate that for lower background
temperature ($T_{b}$=1 keV), the contribution of electron is
dominant (Fig. (3)). By increasing the background temperature, the
contribution of ions increases (Fig. (4)), specially for $\Delta T
\leq 1.5$ MeV it is predominant due to Bragg peak [21]. In Fig. (5)
it can be seen that for $T_{b}$=100 keV the contribution of ions is
dominant over the entire range of temperature difference.

\section{The Energy Exchange Rate Due to Collective Interactions}
The Lenard-Balescu equation for the case of interest in which each
background plasma species $b$ is in thermal equilibrium by itself,
is described by a Maxwell-Boltzmann distribution function (Eq. (3))
[22],
\begin{equation}
\frac{d\varepsilon^{\textrm{LB}}_{pb}}{dt}=\int\frac{d^{3}p_{p}}{(2\pi\hbar)^{3}}\frac{p^{2}_{p}}{2m_{p}}
\nabla_{\textbf{p}_{b}}.\textbf{L}_{pb},
\end{equation}
where $\textbf{L}_{pb}$ is the Lenard-Balescu Kernel,
\begin{equation}
\textbf{L}_{pb}=\int\frac{d^{3}k}{(2\pi)^{3}}\pi\textbf{k}
\left|\frac{4\pi
Z_{p}eZ_{b}e}{k^{2}\varepsilon(k,v_{p}.\textbf{k})}\right|^{2}I^{\textrm{LB}}_{pb}(v_{p})
,
\end{equation}
In Eq. (20), $\varepsilon(k,v_{p}.\textbf{k})$ is the dielectric
function that is given by [23].
\begin{equation}
\varepsilon(\textbf{k},\omega)=1+\Sigma_{c}\frac{4\pi
(Z_{c}e)^{2}}{k^{2}} \int\frac{d^{3}p_{c}}{(2\pi
\hbar)^{3}}\frac{1}{\omega-\textbf{k}.\textbf{v}_{c}+\textrm{i}\eta}
\textbf{k}.\nabla_{p_{c}}f_{c}\left(\textbf{p}_{c}\right) ,
\end{equation}
where the prescription $\eta \rightarrow 0^{+}$ is implicit and
defines the correct retarded response. The structure of the
dielectric function can be simplified as,
\begin{equation}
    k^{2}\varepsilon(\textbf{k},\omega)=k^{2}+F(\omega),
\end{equation}
where
\begin{equation}
    F(\omega)=F_{p}(\omega)+F_{b}(\omega),
\end{equation}
The F functions appear in the form of a dispersion relation,
\begin{eqnarray}
\left\{ \begin{array}{ll} F_{p}(\omega)=-4\pi (Z_{p}e)^{2}
\int\frac{d^{3}p_{p}}{(2\pi \hbar)^{3}}f_{p}(\textbf{p}_{p})
\left\{\frac{1}{p^{2}_{p}}+\frac{2\alpha}{\Delta m_{p}}
 \left(\frac{p^{2}_{p}}{2\Delta m_{p}}-\frac{E_{0}}{\Delta}\right)\right\}
 \frac{\textbf{k}.\textbf{p}_{p}}{\omega-\textbf{k}.\textbf{v}_{p}+\textrm{i}\eta}\\
F_{b}(\omega)=-\sum_{b}\frac{4\pi (Z_{b}e)^{2}\beta_{b}}{m_{b}}
\int\frac{d^{3}p_{b}}{(2\pi \hbar)^{3}}f_{b}(\textbf{p}_{b})
 \frac{\textbf{k}.\textbf{p}_{b}}{\omega-\textbf{k}.\textbf{v}_{b}+\textrm{i}\eta},
                    \end{array}
                    \right.
\end{eqnarray}
These functions can be written as dispersion relations,
\begin{eqnarray}
\left\{ \begin{array}{ll}
F_{p}(\omega)=-\int_{0}^{\infty}dv \frac{\rho_{p}(v)}{\frac{\omega}{k}-v+\textrm{i}\eta}\\
F_{b}(\omega)=-\int_{-\infty}^{\infty}dv
\frac{\rho_{total}(v)}{\frac{\omega}{k}-v+\textrm{i}\eta},
                    \end{array}
                    \right.
\end{eqnarray}
with the spectral weight,
\begin{eqnarray}
\left\{ \begin{array}{ll}
\rho_{p}(v)=\frac{k^{2}_{p}\sqrt{\alpha}}{2\beta_{p}\Delta}
\left\{\frac{1}{v}+\frac{2\alpha
m_{p}v}{\Delta}\left(\frac{m_{p}v^{2}}{2\Delta}-\frac{E_{0}}{\Delta}\right)\right\}
\textrm{exp}\left\{-\alpha \left(\frac{m_{p}v^{2}}{2\Delta}-\frac{E_{0}}{\Delta}\right)^{2}\right\}\\
\rho_{total}(v)=\sum_{b}\rho_{b}(v),
                    \end{array}
                    \right.
\end{eqnarray}
where $\rho_{b}(v)$ is the contribution of species b to the total
spectral weight,
\begin{equation}
\rho_{b}(v)=k^{2}_{b}v\sqrt{\frac{\beta_{b}m_{b}}{2\pi}}\textrm{exp}\left\{-\frac{1}{2}\beta_{b}m_{b}v^{2}\right\}
\end{equation}
and $k^{2}_{c}$ is the contribution of c particle with charge
$Z_{c}e$ to the squared Debye wave number
\begin{equation}
    k^{2}_{c}=4\pi \beta_{c}Z^{2}_{c}e^{2}n_{c},
\end{equation}

The integral $I^{\textrm{LB}}_{pb}(v_{p})$ in Eq. (20) is obtained
as,
\begin{equation}
I^{\textrm{LB}}_{pb}(v_{p})=\int\frac{d^{3}p_{p}}{(2\pi\hbar)^{3}}
\delta(\textbf{k}.\textbf{v}_{p}-\textbf{k}.\textbf{v}_{b})
\textbf{k}.\left[\nabla_{p_{b}}-\nabla_{p_{p}}\right]
f_{p}\left(\textbf{p}_{p}\right)f_{b}\left(\textbf{p}_{b}\right) ,
\end{equation}
the gradient of distribution functions are,
\begin{eqnarray}
\left\{ \begin{array}{ll}
 \nabla_{p_{p}}f_{p}\left(\textbf{p}_{p}\right)=-\left\{\frac{1}{p^{2}_{p}}+\frac{2\alpha}{\Delta m_{p}}
 \left(\frac{p^{2}_{p}}{2\Delta m_{p}}-\frac{E_{0}}{\Delta}\right)\right\}
 \textbf{p}_{p}f_{p}\left(\textbf{p}_{p}\right)\\
 \nabla_{p_{b}}f_{b}\left(\textbf{p}_{b}\right)=-\frac{\beta_{b}}{m_{b}}\textbf{p}_{b}f_{b}\left(\textbf{p}_{b}\right),
                    \end{array}
                    \right.
\end{eqnarray}
replacing in Eq. (29) yield,
\begin{eqnarray}
I^{\textrm{LB}}_{pb}(\textbf{v}_{p})&=&\int\frac{d^{3}p_{p}}{(2\pi\hbar)^{3}}
\delta(\textbf{k}.\textbf{v}_{p}-\textbf{k}.\textbf{v}_{b})\nonumber\\
&&\textbf{k}.\left\{\frac{\beta_{b}\textbf{p}_{b}}{m_{b}}+\textbf{p}_{p}
\left[\frac{1}{p^{2}_{p}}+\frac{2\alpha}{\Delta
m_{p}}\left(\frac{p^{2}_{p}}{2\Delta
m_{p}}-\frac{E_{0}}{\Delta}\right)\right]\right\}
f_{p}\left(\textbf{p}_{p}\right)f_{b}\left(\textbf{p}_{b}\right),
\end{eqnarray}
by separating the parallel and vertical elements of velocities with
$k$, and delta function properties, we have,
\begin{eqnarray}
\left\{ \begin{array}{ll}
\delta(\textbf{k}.\textbf{v}_{p}-\textbf{k}.\textbf{v}_{b})=
\delta\left(k\left[v_{p\|}-v_{b\|}\right]\right)=
\frac{1}{k}\delta\left(v_{p\|}-v_{b\|}\right)=\frac{1}{k}\delta\left(v_{b\|}-v_{p\|}\right)\\
\int
dv_{b\|}\delta\left(v_{b\|}-v_{p\|}\right)f(v_{b\|})=f(v_{p\|}),
                    \end{array}
                    \right.
\end{eqnarray}
the integral Eq. (31) is obtained as,
\begin{equation}
I^{\textrm{LB}}_{pb}(\textbf{v}_{p})=-\frac{\rho_{b}(v_{p\|})}{4\pi
\beta_{b} Z^{2}_{b}e^{2}} \left\{\beta_{b}-m_{p}
\left[\frac{1}{p^{2}_{p}}+\frac{2\alpha}{\Delta
m_{p}}\left(\frac{p^{2}_{p}}{2\Delta
m_{p}}-\frac{E_{0}}{\Delta}\right)\right]\right\}
f_{p}\left(\textbf{p}_{p}\right),
\end{equation}
By using the divergence relation,
\begin{equation}
\frac{p^{2}_{p}}{2m_{p}}\nabla_{\textbf{p}_{p}}\textbf{.L}_{pb}=
\nabla_{\textbf{p}_{p}}.\left(\frac{p^{2}_{p}}{2m_{p}}\textbf{L}_{pb}\right)
-\textbf{L}_{pb}.\nabla_{\textbf{p}_{p}}\left(\frac{p^{2}_{p}}{2m_{p}}\right)
,
\end{equation}
and the gradient of kinetic energy of projectile,
\begin{equation}
\nabla_{\textbf{p}_{p}}\left(\frac{p^{2}_{p}}{2m_{p}}\right)=
\frac{\textbf{p}_{p}}{m_{p}}=\textbf{v}_{p} ,
\end{equation}
the Eq. (19) changes as,
\begin{equation}
\frac{d\varepsilon^{\textrm{LB}}_{pb}}{dt}=\int\frac{d^{3}p_{p}}{(2\pi\hbar)^{3}}\nabla_{\textbf{p}_{b}}.
\left(\frac{p^{2}_{p}}{2m_{p}}\textbf{L}_{pb}\right)-\int\frac{d^{3}p_{p}}{(2\pi\hbar)^{3}}\textbf{L}_{pb}.\textbf{v}_{p}=
-\int\frac{d^{3}p_{p}}{(2\pi\hbar)^{3}}\textbf{L}_{pb}.\textbf{v}_{p}
,
\end{equation}
putting the Eqs. (20) and (33) in this equation yields,
\begin{eqnarray}
\frac{d\varepsilon^{\textrm{LB}}_{pb}}{dt}&=&\int\frac{d^{3}p_{p}}{(2\pi
\hbar)^{3}}
\int\frac{d^{3}k}{(2\pi)^{3}}\pi\textbf{k}.\textbf{v}_{p}
\left|\frac{4\pi Z_{p}eZ_{b}e}{k^{2}\varepsilon(k,v_{p}.\textbf{k})}\right|^{2}\nonumber\\
&&\times \frac{\rho_{b}(v_{p\|})}{4\pi \beta_{b} Z^{2}_{b}e^{2}}
\left\{\beta_{b}-m_{p}
\left[\frac{1}{p^{2}_{p}}+\frac{2\alpha}{\Delta
m_{p}}\left(\frac{p^{2}_{p}}{2\Delta
m_{p}}-\frac{E_{0}}{\Delta}\right)\right]\right\}
f_{p}\left(\textbf{p}_{p}\right),
\end{eqnarray}
separating the parallel and vertical elements of velocities with $k$
and Eq. (22), with a little calculation the temperature
equilibration rate is obtained as,
\begin{eqnarray}
&&\frac{d\varepsilon^{\textrm{LB}}_{pb}}{dt}=
\frac{n_{p}\alpha^{1/2}Z^{2}_{p}e^{2}}{\Delta
m_{p}\pi^{1/2}\beta_{b}}
\int^{\infty}_{0}\frac{dkk^{3}}{\left|k^{2}+F(kp_{p}/m_{p})\right|^{2}}\int^{\infty}_{0}
dp_{p}
(p_{p}/m_{p})\rho_{b}(p_{p}/m_{p}) \nonumber\\
&&\times \left\{p_{p}\beta_{b}-m_{p}
\left[\frac{1}{p_{p}}+\frac{2\alpha p_{p}}{\Delta
m_{p}}\left(\frac{p^{2}_{p}}{2\Delta
m_{p}}-\frac{E_{0}}{\Delta}\right)\right]\right\} \exp\left\{-\alpha
\left(\frac{p^{2}_{p}}{2\Delta
m_{p}}-\frac{E_{0}}{\Delta}\right)^{2}\right\},
\end{eqnarray}
For shorthand, this equation is wrtten as,
\begin{equation}
\frac{d\varepsilon^{\textrm{LB}}_{pb}}{dt}=
\frac{n_{p}\alpha^{1/2}Z^{2}_{p}e^{2}}{\Delta
m_{p}\pi^{1/2}\beta_{b}} \int^{\infty}_{0} dp_{p}G(p_{p})Q(p_{p})
(p_{p}/m_{p})\rho_{b}(p_{p}/m_{p}),
\end{equation}
where the function $Q(p_{p})$ is,
\begin{equation}
Q(p_{p})= \left\{p_{p}\beta_{b}-m_{p}
\left[\frac{1}{p_{p}}+\frac{2\alpha p_{p}}{\Delta
m_{p}}\left(\frac{p^{2}_{p}}{2\Delta
m_{p}}-\frac{E_{0}}{\Delta}\right)\right]\right\} \exp\left\{-\alpha
\left(\frac{p^{2}_{p}}{2\Delta
m_{p}}-\frac{E_{0}}{\Delta}\right)^{2}\right\},
\end{equation}
and the function $G(p_{p})$ is defined and solved as,
\begin{eqnarray}
&&G(p_{p})=
\int^{\infty}_{0}\frac{dkk^{3}}{\left|k^{2}+F(kp_{p}/m_{p})\right|^{2}}= \nonumber\\
&&\frac{y(kp_{p}/m_{p})\ln\left[x^{2}(kp_{p}/m_{p})+y^{2}(kp_{p}/m_{p})\right]+x(kp_{p}/m_{p})
\arctan\left[\frac{y(kp_{p}/m_{p})}{x(kp_{p}/m_{p})}\right]}{-4y(kp_{p}/m_{p})},
\end{eqnarray}
where $x(\omega)$ and $y(\omega)$ are the real and imaginary of
$F(\omega)$ function,
\begin{eqnarray}
\left\{ \begin{array}{ll}
x(\omega)=\Re[F(\omega)]\\
y(\omega)=\Im[F(\omega)],
                    \end{array}
                    \right.
\end{eqnarray}
The results of solving Eq. (39) are drawn in Figs. (6), (7) and (8)
for background temperatures $T_{b}$=1 keV, 10 keV and 100 keV,
respectively. The ions and electron contributions are denoted by the
dashed and dotted lines, respectively, and the solid line represents
the sum of ions and electron contributions. In general, the results
show that the contribution of ions increases for higher background
temperatures.
\section{Total Temperature Equilibration Rate}
The total temperature equilibration rate is obtained by adding
contributions of the binary collisions (Eq. (16)) and collective
interactions (Eq. (39)) as,
\begin{equation}
\frac{d\varepsilon_{pb}}{dt}=\frac{d\varepsilon^{\textrm{B}}_{pb}}{dt}+
\frac{d\varepsilon^{\textrm{LB}}_{pb}}{dt},
\end{equation}
The temperature equilibration rate (Eq. (43)) of quasi-monoenergetic
deuteron beam with an equimolar deuterium-tritium plasma versus
temperature difference $\Delta T$ between projectile and background
particles, for background temperatures $T_{b}$=1 keV, 10 keV and 100
keV are depicted in Figs. (9), (10) and (11), respectively.
 The highest projectile temperature is chosen $T_{D}$=10 MeV and the background density is taken $\rho_{b}=300
~\textrm{gcm}^{-3}$. In these figures, the contributions of the
binary collisions and the collective interactions are denoted by the
dashed and dotted lines, respectively, and the solid line represents
the total temperature equilibration rate. The results predicate
reduction of the total temperature equilibration rate by increasing
the background temperature, generally. Furthermore, independent of
background temperature,
 the temperature equilibration rate due to binary collisions has greater contribution than collective interactions.
In Fig. (9) ($T_{b}$=1 keV), the Bragg peak is not visible because
the electron contribution dominates for fewer background
temperatures. In contrast, for greater background temperatures
(Figs. (10 and (11)) the Bragg peak which is quite obvious at the
lower temperature differences, is a property of Coulomb collisions
between ions. Also, at the higher temperature differences,
temperature equilibration rate increases due to the electron
contribution and nuclear elastic scattering in the binary
collisions.
\section{Conclusion}
The thermal equilibrium between charged particles is crucial to understand
the overall energy balance in a fusion plasma, where
the ignition and burn of the plasma are strongly temperature
dependent. Since quasi-monoenergetic beams are appropriate choices
for ignitor in fast ignition, we calculate their temperature
equilibration rate in the fusion plasma. According to the obtained
results, the temperature equilibration rate increases highly by increasing projectile and background particles temperature
difference, $\Delta T$, for high background particles temperature.
In such condition, the major contribution of exchange of energy is
related to the binary collisions between projectile ions and
background electrons. Increasing the background temperature, the
energy exchange rate is reduced gradually. In this case, the major
contribution of exchange of energy is related to the interactions
between the projectile and the background ions. Figs. (9, 10 and
11) show that, at low as well as high temperatures of the
fuel,($T_{b}$=1 keV, 10 keV and 100 keV), the contribution of
collective interactions is lower than the contribution of  binary
collisions in the energy exchange rate.


\newpage
\begin{figure}
\centerline{ \epsfxsize=15cm \epsffile{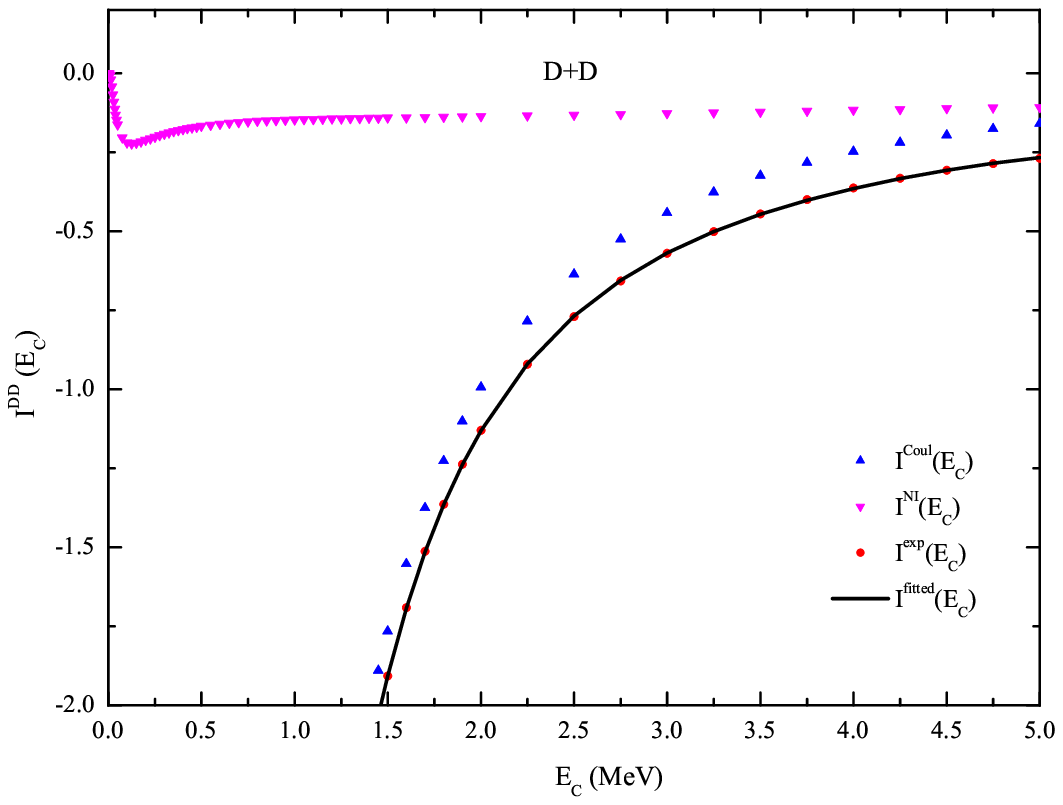}} \caption{The
integral $I_{1}(E_{C})$ (Eq. (15)) versus
 the total energy in the center of mass coordinate $E_{C}$
 for D+D scattering. The contributions of Coulomb, nuclear and experimental data are depicted, separately.}
\end{figure}
\newpage
\begin{figure}
\centerline{ \epsfxsize=15cm \epsffile{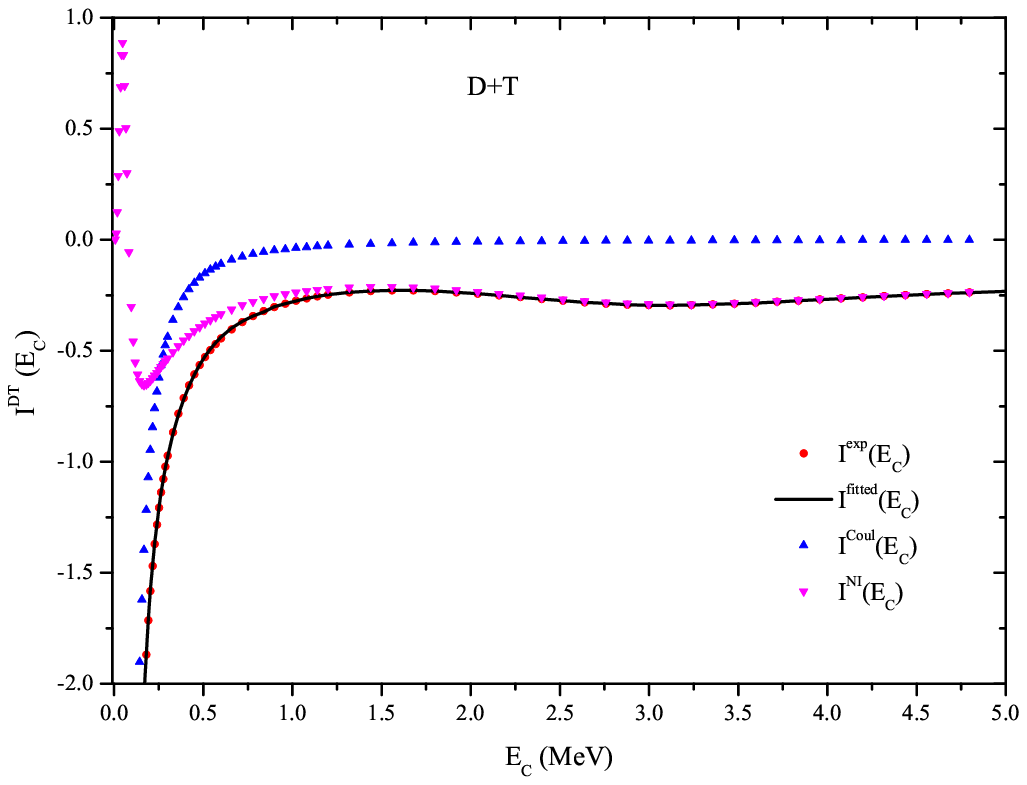}} \caption{The
integral $I_{1}(E_{C})$ (Eq. (15)) versus
 the total energy in the center of mass coordinate $E_{C}$
 for D+T scattering. The contributions of Coulomb, nuclear and experimental data are depicted, separately.}
\end{figure}

\newpage
\begin{figure}
\centerline{ \epsfxsize=15cm \epsffile{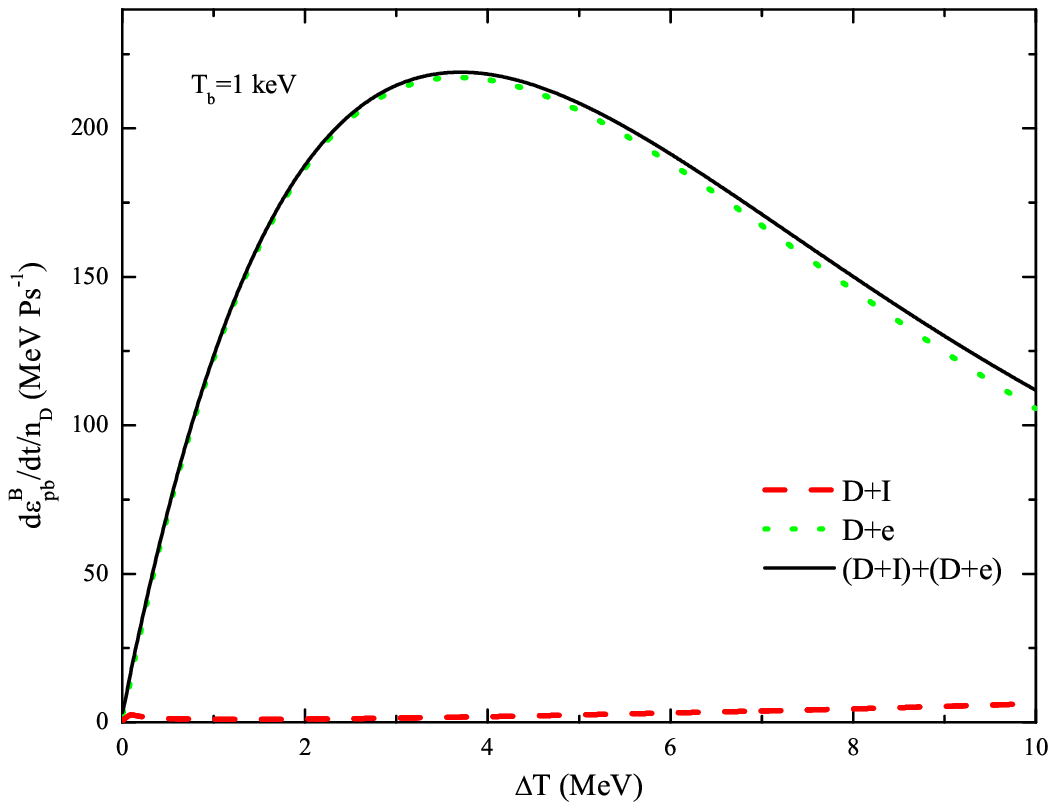}} \caption{The
temperature equilibration rate due to binary collisions (Eq.(16))
versus projectile and background particles temperature difference,
$\Delta T$, for background temperature $T_{b}$=1 keV and density
$\rho_{b}=300~{\textrm{gcm}^{-3}}$. The dashed and dotted lines
denote ions and electron contributions, respectively and solid line
is the total temperature equilibration rate.}
\end{figure}
\newpage
\begin{figure}
\centerline{ \epsfxsize=15cm \epsffile{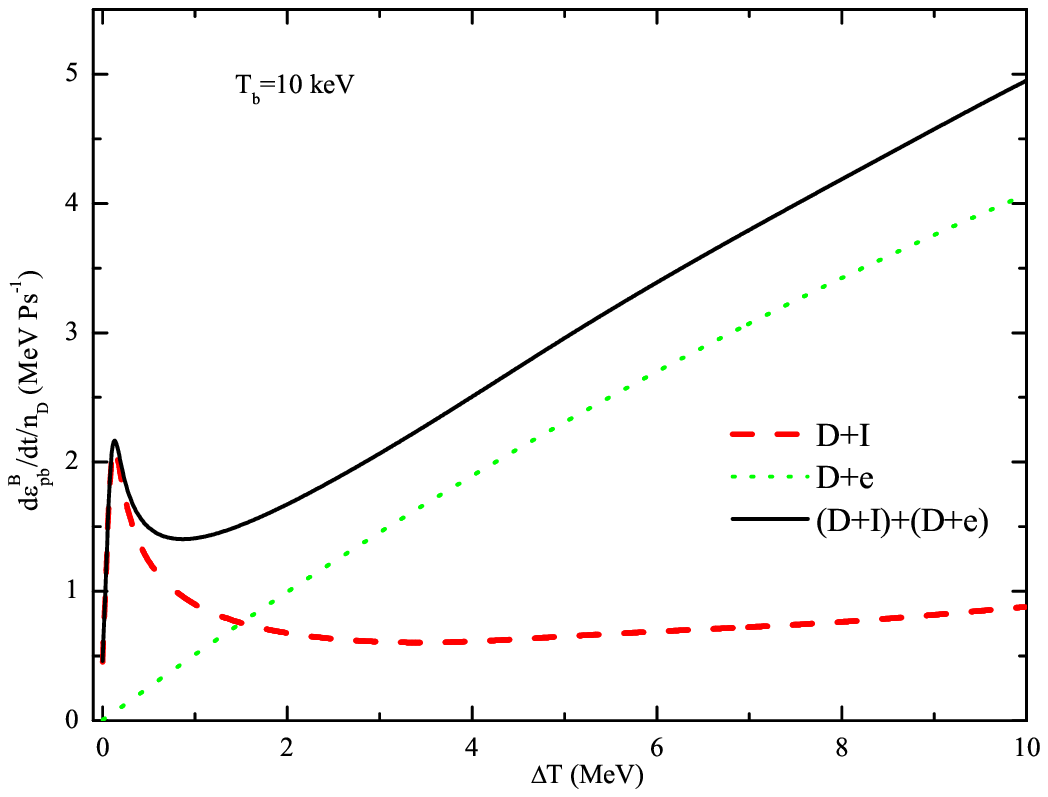}} \caption{The
temperature equilibration rate due to binary collisions (Eq.(16))
versus projectile and background particles temperature difference,
$\Delta T$, for background temperature $T_{b}$=10 keV and density
$\rho_{b}=300~{\textrm{gcm}^{-3}}$. The dashed and dotted lines
denote ions and electron contributions, respectively and solid line
is the total temperature equilibration rate.}
\end{figure}
\newpage
\begin{figure}
\centerline{ \epsfxsize=15cm \epsffile{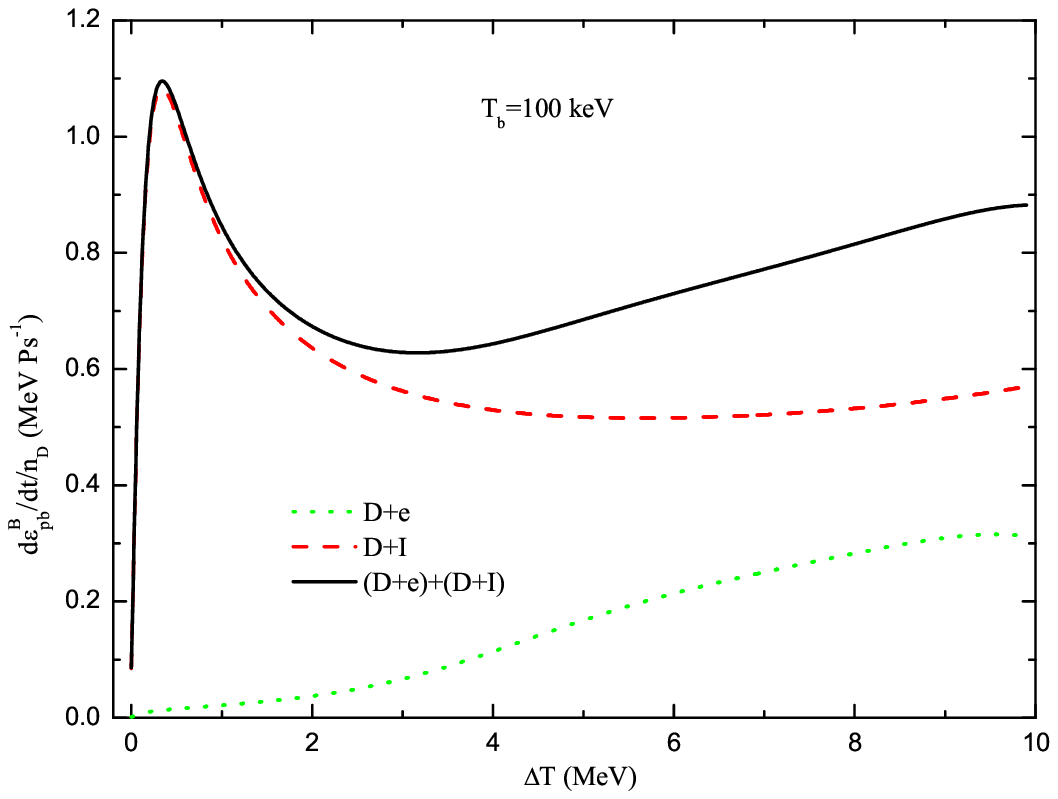}} \caption{The
temperature equilibration rate due to binary collisions (Eq.(16))
versus projectile and background particles temperature difference,
$\Delta T$, for background temperature $T_{b}$=100 keV and density
$\rho_{b}=300~{\textrm{gcm}^{-3}}$. The dashed and dotted lines
denote ions and electron contributions, respectively and solid line
is the total temperature equilibration rate.}
\end{figure}
\newpage
\begin{figure}
\centerline{ \epsfxsize=15cm \epsffile{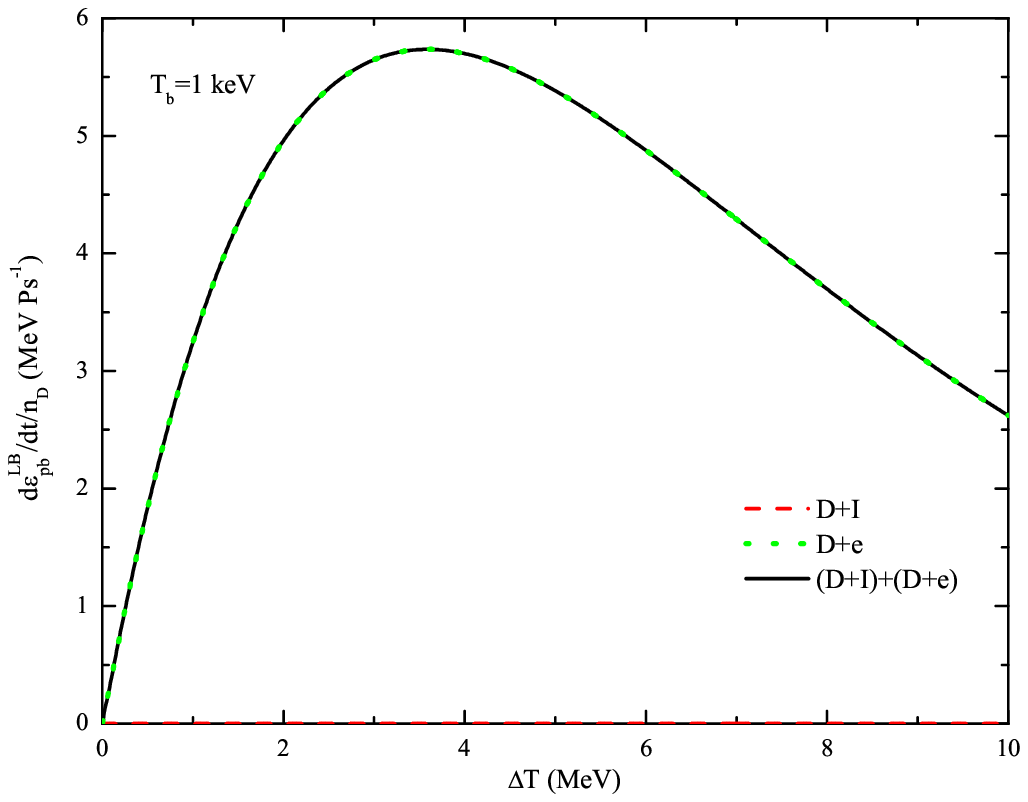}} \caption{The
temperature equilibration rate due to collective interactions
(Eq.(39)) versus projectile and background particles temperature
difference, $\Delta T$, for background temperature $T_{b}$=1 keV and
density $\rho_{b}=300~{\textrm{gcm}^{-3}}$. The dashed and dotted
lines denote ions and electron contributions, respectively and solid
line is the total temperature equilibration rate.}
\end{figure}
\newpage
\begin{figure}
\centerline{ \epsfxsize=15cm \epsffile{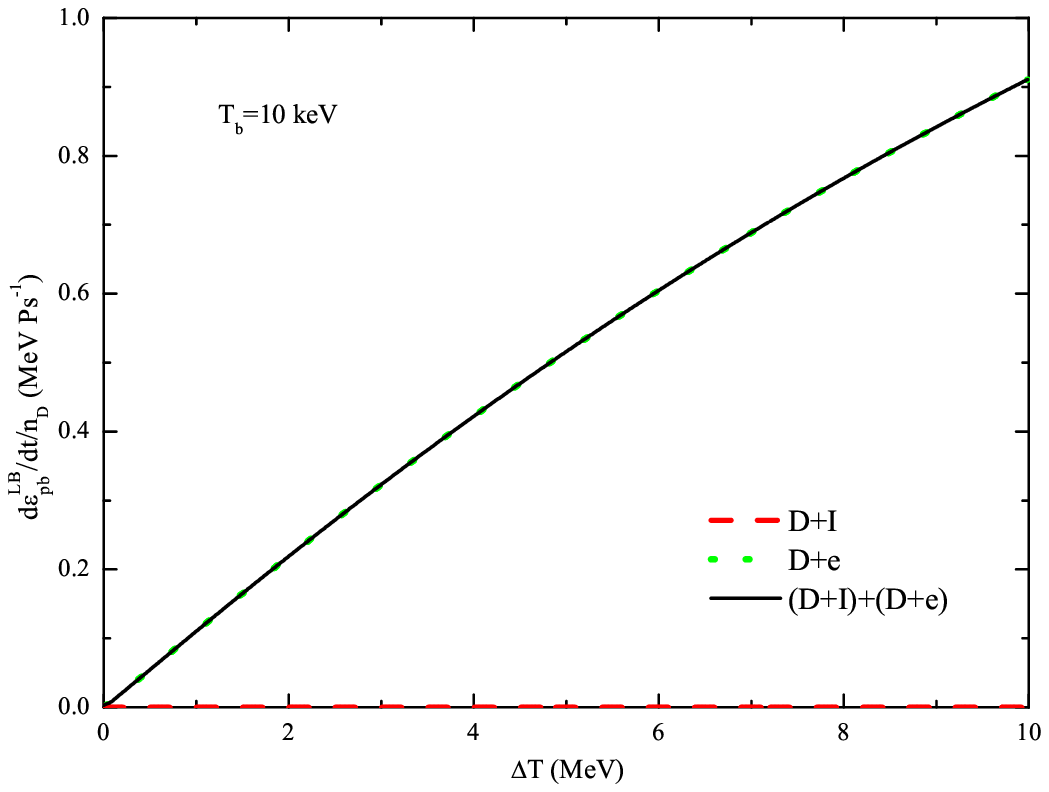}} \caption{The
temperature equilibration rate due to collective interactions
(Eq.(39)) versus projectile and background particles temperature
difference, $\Delta T$, for background temperature $T_{b}$=10 keV
and density $\rho_{b}=300~{\textrm{gcm}^{-3}}$. The dashed and
dotted lines denote ions and electron contributions, respectively
and solid line is the total temperature equilibration rate.}
\end{figure}
\newpage
\begin{figure}
\centerline{ \epsfxsize=15cm \epsffile{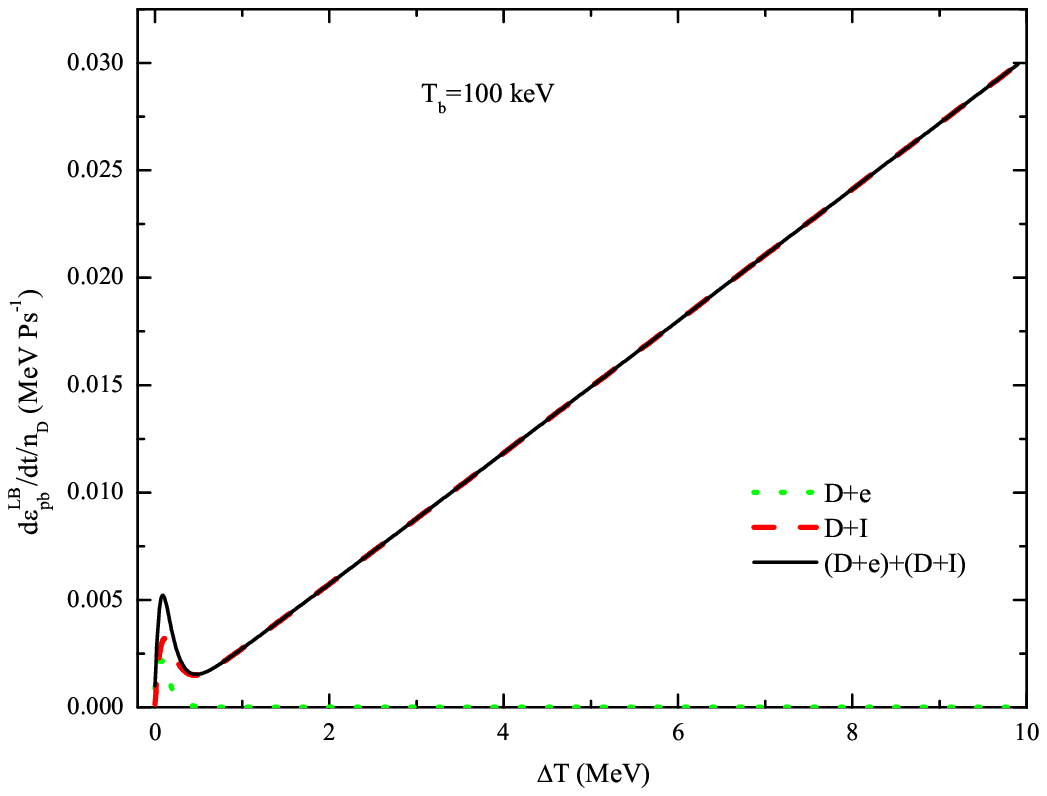}} \caption{The
temperature equilibration rate due to collective interactions
(Eq.(39)) versus projectile and background particles temperature
difference, $\Delta T$, for background temperature $T_{b}$=100 keV
and density $\rho_{b}=300~{\textrm{gcm}^{-3}}$. The dashed and
dotted lines denote ions and electron contributions, respectively
and solid line is the total temperature equilibration rate.}
\end{figure}
\newpage
\begin{figure}
\centerline{ \epsfxsize=15cm \epsffile{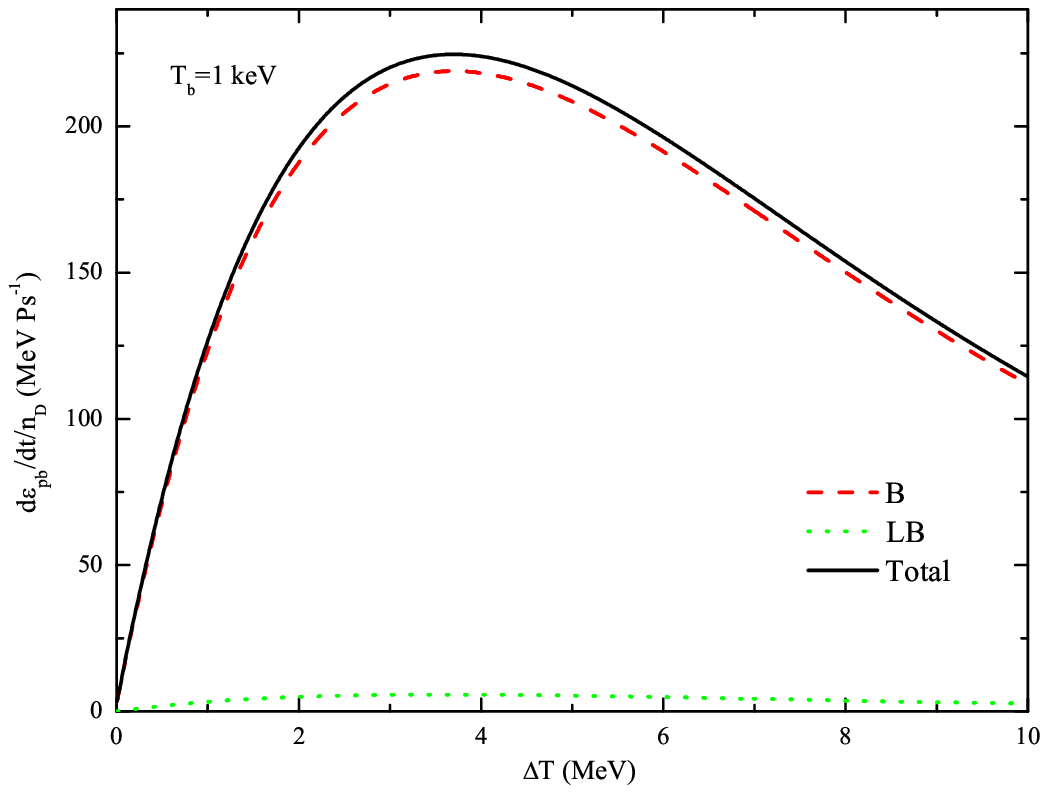}} \caption{The
temperature equilibration rate versus projectile and background
particles temperature difference, $\Delta T$, for background
temperature $T_{b}$=1 keV and density
$\rho_{b}=300~{\textrm{gcm}^{-3}}$. The dashed and dotted lines
denote binary collisions (Eq. (16)) and collective interactions (Eq.
(39)), respectively and solid line is the total temperature
equilibration rate (Eq. (43)).}
\end{figure}
\newpage
\begin{figure}
\centerline{\epsfxsize=15cm \epsffile{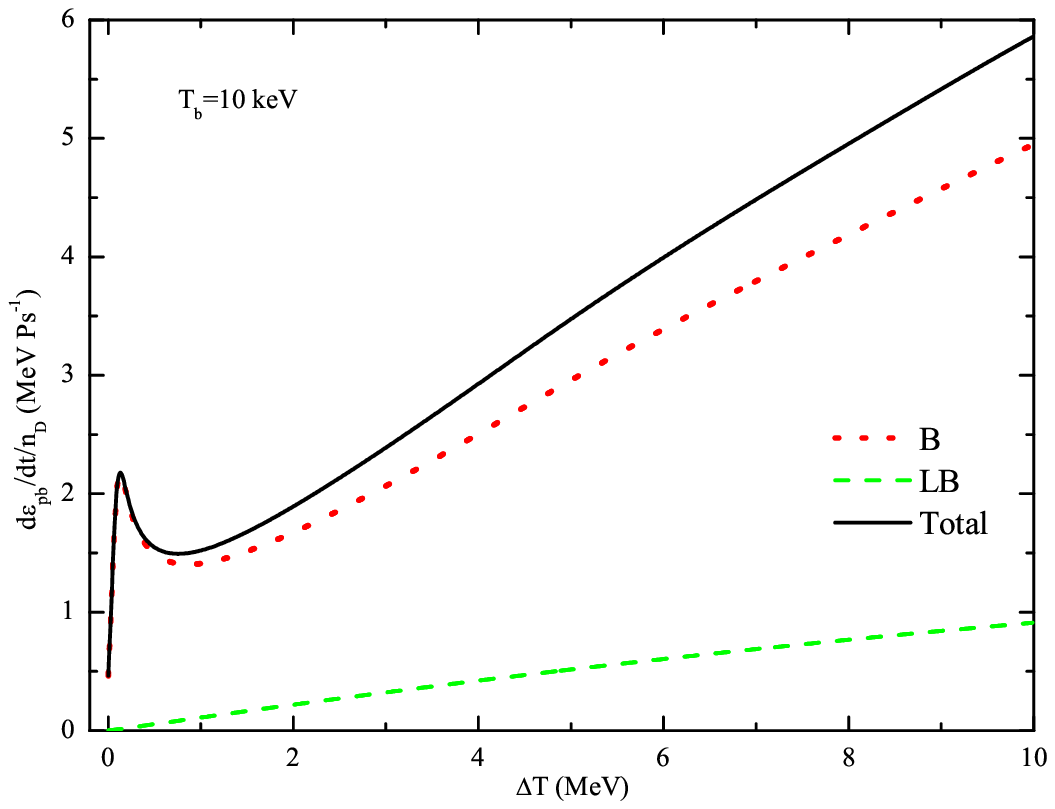}} \caption{The
temperature equilibration rate versus projectile and background
particles temperature difference, $\Delta T$, for background
temperature $T_{b}$=10 keV and density
$\rho_{b}=300~{\textrm{gcm}^{-3}}$. The dashed and dotted lines
denote binary collisions (Eq. (16)) and collective interactions (Eq.
(39)), respectively and solid line is the total temperature
equilibration rate (Eq. (43)).}
\end{figure}
\newpage
\begin{figure}
\centerline{ \epsfxsize=15cm \epsffile{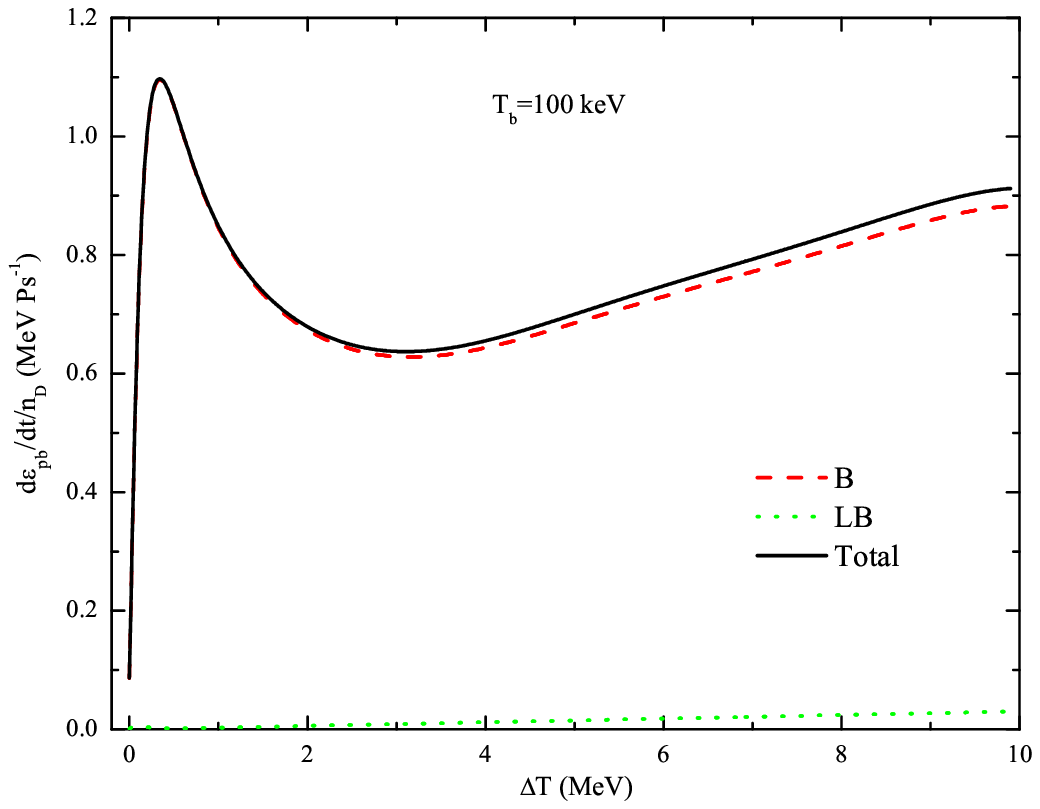}} \caption{The
temperature equilibration rate versus projectile and background
particles temperature difference, $\Delta T$, for background
temperature $T_{b}$=100 keV and density
$\rho_{b}=300~{\textrm{gcm}^{-3}}$. The dashed and dotted lines
denote binary collisions (Eq. (16)) and collective interactions (Eq.
(39)), respectively and solid line is the total temperature
equilibration rate (Eq. (43)).}
\end{figure}
\newpage
\end{document}